\documentclass{TTP_DSL2006}
\usepackage{floatflt}
\usepackage[dvipsone]{graphicx}
\usepackage[intlimits]{amsmath}
\usepackage{amssymb}
\usepackage{empheq}
\usepackage{exscale}
\begin{document}
\title{Beyond the Point Ps Approximation}
%\title{¬ыход за рамки приближени€ атома позитрони€ точечной частицей в позитронной аннигил€ционной спектроскопии}
\author{Sergey V. Stepanov\inst{1}$^{,\rm{a}}$\textbf{,} Dmitry S. Zvezhinskiy\inst{1}\textbf{,}
		Vsevolod M. Byakov\inst{1,2}}
 \institute{$^1$Institute of Theoretical and Experimental Physics, B.Cheremushkinskaya 25, 117218 Moscow, Russia\\
 $^2$D.Mendeleyev University of Chemical Technology, Miusskaya sq., 9, Moscow 125047, Russia}
 \maketitle
%\vspace{-6mm}
\sffamily
 \begin{center} $^{a}$stepanov@itep.ru \end{center}

\vspace{2mm} \hspace{-7.7mm} \normalsize \textbf{Keywords:}
pick-off annihilation, positronium bubble model, non-point Ps, contact density\\

\vspace{-2mm} \hspace{-7.7mm} \rmfamily \noindent \textbf{Abstract.} In application to positron annihilation
spectroscopy, Ps atom is considered not as a point particle, but as a finite size e$^+$e$^-$ pair localized in
a bubble-state in a medium. Variation of the internal Coulombic e$^+$-e$^-$ attraction vs. the bubble radius is
estimated.

\subsection{Introduction}

Typical lifetimes (up to annihilation) of a para-positronium atom(p-Ps; spin = 0)\footnote{Ps is a bound state
of e$^+$ and e$^-$.} in condensed medium are about 130-180 ps. They are close to the p-Ps lifetime in vacuum (125 ps). The ortho-positronium lifetime in a medium is considerably shorter (about 100 times; some ns) in comparison with that in vacuum. This is due to the so-called pick-off process -- prompt 2$\gamma$-annihilation of the e$^+$, composing Ps atom, with one of the nearest e$^-$ of surrounding molecules, whose spin is antiparallel to the e$^+$ spin. Just this property turns Ps into a nanoscale structural probe of matter. The theoretical task consists in calculating the pick-off annihilation rate $\lambda_{po}$, i.e. in relating $\lambda_{po}$ with such properties of the medium like surface tension, viscosity, external pressure and size of the Ps trap.

Originally, to explain the unexpectedly long lifetime of the ortho-Ps atom in liquid helium R.Ferrel \cite{Fer56_57} suggested that the Ps atom forms a nanobubble around itself. This is caused by a strong exchange repulsion between the o-Ps electron and electrons of the surrounding He atoms. Ferrel approximated this repulsion by a spherically symmetric potential barrier of radius $R_\infty$. To estimate the equilibrium radius of the Ps bubble he minimized the sum of the Ps energy in a spherically symmetric potential well, i.e. $\pi^2\hbar^2/4mR_\infty^2=\frac{\rm Ry}{2} (\pi a_B/R_\infty)^2$, Ry=13.6 eV, and the surface energy, $4\pi R_\infty^2 \sigma$, where $\sigma$ is the macroscopic surface tension coefficient. The following relationship is hereby obtained for the equilibrium radius of the bubble:
\begin{equation}\label{9_2}
\frac{\pi^2 a_B^2}{R_\infty^2} {\rm Ry} + 4\pi R_\infty^2 \sigma ~\leftrightarrow  ~ \hbox{min over} ~R_\infty
 ~~~\Rightarrow ~~~
R_\infty = a_B \left( \frac{\pi {\rm Ry}}{8\sigma a_B^2} \right)^{1/4}.
\end{equation}
%{\red It came out that $R_\infty$ in He is very large: almost 20 times as large then the radius of He atom. /Why this particular case is necessary here?/}

\subsection{The Tao-Eldrup model}

\begin{figure}[t!]
\centering
\begin{minipage}[h]{80mm}
  \includegraphics[width=80mm]{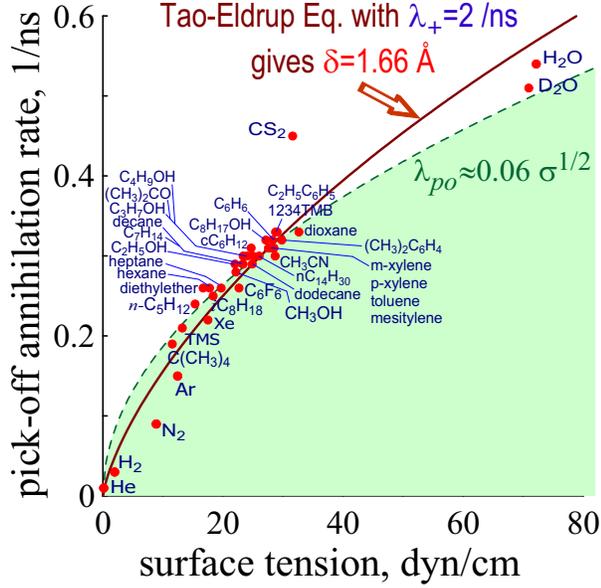}
\end{minipage}
~~~
\begin{minipage}[h]{80mm}
\caption{Dependence of the pick-off annihilation rates \cite{Ste02JSC} vs. surface tension in different liquids.
Solid curve shows the correlation given by the Tao-Eldrup at $\lambda_+ = 2$ ns$^{-1}$ and optimal value $\delta = 1.66$ \AA\ (obtained from fitting of these data by means of Eq. (\ref{9_5})). Dashed curve illustrates simplest approximation $\lambda_{po} \propto \sigma^{1/2}$.} \label{Lpo_sig}
\end{minipage}
\end{figure}

Ferrel's idea got further development in the studies of Tao \cite{Tao72} and Eldrup et al. \cite{Eld81}. They considered the Ps atom as a point particle in a liquid, i.e. in a structureless continuum, Fig. \ref{F_TF}. The repulsive Ps-liquid interaction was approximated by a rectangular infinitely deep spherically symmetric potential well of radius $R_{\infty}$. In such a well, the wave function of a point particle has the following standard expression:
\begin{equation}\label{9_3}
\Psi (0\le r \le R_\infty) = \frac{\sin(\pi r/R_\infty)}{\sqrt{2\pi R_\infty} ~ r},
 \qquad
\Psi (r\ge R_\infty) = 0.
\end{equation}
Here, $r$ is the Ps center-of-mass coordinate. Because the Ps wave function equals to zero at the bubble radius (and outside), there is no e$^+$ overlapping with outer electrons of a medium. So, pick-off annihilation is absent. To overcome this difficulty it was postulated that molecular electrons, which form a ``wall'' of the Ps bubble, may penetrate inside the potential well. This results in the appearance of a surface layer of thickness $\delta = R_{\infty}-R$ having the same average electron density as in the bulk. As a result, the pick-off annihilation rate $\lambda_{po}$ becomes non-zero. It is proportional to the e$^+$ overlapping integral with the electrons inside the bubble:
\begin{equation}\label{9_5}
\lambda_{po} = \lambda_+ P_R, \qquad P_R = \int_{R}^{R_{\infty}} |\Psi(r)|^2 4\pi r^2 dr
%= 1-{R \over R_\infty} + { \sin (2\pi R / R_\infty) \over 2\pi }=
=\frac{\delta}{R_\infty} - \frac{\sin (2\pi \delta / R_\infty)}{2\pi }.
\end{equation}
This is the well-known Tao-Eldrup formula. Here, $\lambda_+\approx 2$ ns$^{-1}$ is the e$^+$ annihilation rate in an unperturbed medium (it is proportional to Dirac's 2$\gamma$-annihilation cross-section and the number density of valence electrons). The thickness $\delta$ of the electron layer is an empirical parameter, which may have different values in various media. Substituting Eq. (\ref{9_2}) for $R_\infty$ into Eq. (\ref{9_5}), one obtains the relationship between $\lambda_{po}$ and $\sigma$ with one adjustable parameter, $\delta$. \footnote{If we make use of the relationship $\frac{\sin x}{x} \approx 1-\frac{x}{\pi}$, which is approximately valid for $x < 4$, one may write
$$
\frac{\delta}{R_\infty} - { \sin (2\pi \delta / R_\infty) \over 2\pi } \approx \frac{2\delta^2}{R_\infty^2}.
$$
Thus we obtain $\lambda_{po} \propto \sigma^{1/2}$, which is displayed in Fig. \ref{Lpo_sig}.}
It may be easily obtained by fitting experimental pick-off annihilation rates with the relationship (\ref{9_5}), Fig. \ref{Lpo_sig}. Thus we obtain $\delta \approx 1.66$ \AA. Eq. (\ref{9_5}) with this value of $\delta$ is widely used for recalculation of the observed pick-off annihilation rate into the free volume $4\pi R^3/3$ of the cavity, where Ps atom resides and annihilates.

%ќтметим, что по случайному совпадению энерги€ Ps, выражаема€ формулой (\ref{9_4}) при $R_\infty = \delta$, оказываетс€ равной Ry/2 поскольку $\pi a_B \approx 1.66$ \AA.

%%%%%%%%%%%%%%%%%%%%%%%%%%%%%%%%%%%%%%%%%%%%%%%%%%%%%%%%%%%
%%%%%%%%%%%%%%%%%%%%%%%%%%%%%%%%%%%%%%%%%%%%%%%%%%%%%%%%%%%%

%Ёту же формулу часто используют и дл€ определени€ величины свободного объема в полимерах. ѕри этом, однако,
%следует иметь ввиду, что предсуществующие полости, в которых локализуетс€ позитроний, не €вл€ютс€
%термодинамически равновесными (как в жидкост€х), и параметр $\delta$ может отличатьс€ от значени€ 1.66 \AA.

\subsection{Further development of the Ps bubble models}

Along with the development of the ``infinite potential well'' Ps bubble model, another approach based on the finite potential well approximation was also elaborated \cite{Stew59,Bych61,Roel67,Dau00,Ste02JSC}. However in both approaches, the Ps atom was approximated by a point particle. This leads to a significant simplification, but it is not justified from a physical viewpoint, because:

1) the size of the localized state of Ps (size of the Ps bubble) does not significantly exceed the distance between e$^+$ and e$^-$ in Ps;

2) during the formation of the Ps bubble there is a substantial variation of the Ps internal energy (particularly of the Coulombic attraction of e$^+$ and e$^-$), which is completely ignored in the ``point-like'' Ps models. In a vacuum or in a large bubble, the internal energy of Ps tends to $-{\rm Ry}/2=-6.8$ eV. In a continuous liquid (no bubble) with the high-frequency dielectric permittivity $\varepsilon \approx n^2$ ($n\approx 2$-3 is the refractive index) the energy of the Coulombic attraction between e$^+$ and e$^-$ decreases in absolute value by a factor $\varepsilon^2 \approx 4$-9. The same takes place with the total Ps binding energy, which tends to the value $-{\rm Ry}/2\varepsilon^2 \approx -(1$-1.7) eV (this is a simple consequence of the scaling $e^2 \to e^2/\varepsilon$ of the Schr\"odinger equation for Ps atom). Thus, the change in the Ps internal energy during Ps formation may reach 5 eV. Obviously, this represents an important contribution to the energetics of Ps formation. The aim of the present work is towards a more accurate estimation of this contribution, that has not been done yet.

There is only a small number of papers where the consequences of the finite size of Ps are discussed in application to positron annihilation spectroscopy. To calculate $\lambda_{po}$, the Kolkata group \cite{Dut02} suggested to smear the Ps atom over the relative e$^+$-e$^-$ coordinate exactly in the same way as it is in a vacuum. Such an approach is valid for rather large bubbles. However, they do not discuss the variation of the internal Ps energy.

In \cite{Seeg03} the Ps atom is considered as a finite sized e$^-$e$^+$ pair, but the variation of the Coulombic interaction because of dielectric screening is not discussed. It was assumed that e$^-$ is confined in an infinite potential well and e$^+$ is bound to it by means of the Coulombic attraction. The wave function of the pair was taken as a series of orthogonal polynomials, their weights being determined from a minimization procedure of the total energy of the pair.

\subsection{Hamiltonian of e$^+$e$^-$ pair in a medium}

Let the e$^+$e$^-$ pair (Ps atom) have already formed in a liquid a nanobubble (spherical cavity; Ps bubble) of radius $R$ (the onset of coordinates is taken at the center of the bubble, Fig. \ref{Bub_coord}). Together with the molecules surrounding the e$^+$e$^-$ pair, one has to deal with a quite intricate many-body problem with a complex hamiltonian. We reduce it to the following form:
\begin{equation}\label{2_1}
H \approx -\frac{\hbar^2 (\Delta_+ + \Delta_-)}{2m_e} + U({\bf r}_+) + U({\bf r}_-)
 - U_c({\bf r}_+, {\bf r}_-, R, \varepsilon).
%  \frac{e^2}{\varepsilon({\bf r}_+, {\bf r}_-) |{\bf r}_+ - {\bf r}_-|} +
\end{equation}
%%%%%%%%%%%%%%%%%%%%%%%%%%%%%%%%%%%%%%%%%%%%%%%%%%%%%%%%%%%
\noindent
 \begin{floatingfigure}[h!]{35mm} % add floatflt; l-left,r-right 45mm-width
% \noindent \epsfig{file=./Bubble_coordinates.eps, width=35mm}
  \noindent \includegraphics[width=35mm]{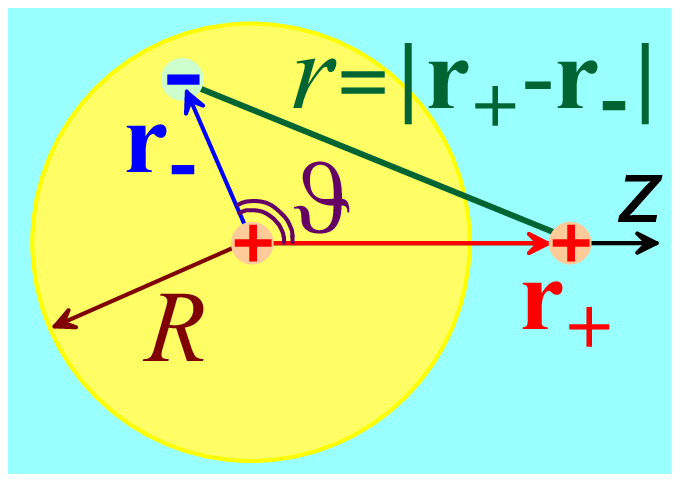}
 \caption{}
  \label{Bub_coord}
 \end{floatingfigure}
%%%%%%%%%%%%%%%%%%%%%%%%%%%%%%%%%%%%%%%%%%%%%%%%%%%%%%%%%%%%
\noindent Terms with Laplacians $\Delta_+$ and $\Delta_-$ over ${\bf r}_+$ and ${\bf r}_-$ (e$^+$ and e$^-$ coordinates) stand for the kinetic energies of the particles. $U({\bf r}_+)$ and $U({\bf r}_-)$ describe the individual interaction of e$^+$ and e$^-$ with the medium. For them we adopt the following approximation:
\begin{equation}\label{2_2}
U(r_+) = \left\{ \begin{tabular}{ll}
						  0 , & $r_+<R$,\\
						 $V_0^+$, & $r_+>R$,\\
						 \end{tabular}    \right.   \qquad
U(r_-) = \left\{ \begin{tabular}{ll}
						  0 , & $r_-<R$,\\
						 $V_0^-$, & $r_->R$.\\
						 \end{tabular}    \right.
\end{equation}
Here, $V_0^+$ and $V_0^-$ are the e$^+$ and e$^-$ work functions, respectively ($V_0$ is a commoner notation for the electron work function). The work function is usually introduced as the energy needed for an excess particle to enter the liquid without any rearrangement of its molecules and to stay there in a delocalized state, having no preferential location in a bulk. One may say that $V_0^+$ and $V_0^-$ are the ground state energies of the quasifree e$^+$ and e$^-$, because their energies at rest after having been removed from the liquid to infinity are defined to be zero.

\begin{table}
\caption{Electron work function for different liquids at room temperature \cite{CRC}}  \label{V0}
\bigskip
 \centering
\begin{tabular}{|lc||lc|}
\hline
 Liquid             & $V_0^-$, eV  & Liquid        & $V_0^-$, eV \\
\hline
 helium; 4.2 K       &         1.3  & benzene        &-0.14 \\
 n-dodecane          &         0.2  & isooctane      &-0.17 \\
 n-decane            &         0.18 & toluene        &-0.22 \\
 n-heptane           &         0.12 & neopentane     &-0.38  \\
 n-hexane            &         0.1  & MeOH, EtOH, PrOH &-0.4\\
 nitrogen; 77.3 K    &         0.05 & xenon; 170 K   &-0.57  \\
 n-pentane, c-hexane &         0.01 & water          &-1.2   \\
 argon; 86.4 K       &         0    &                &       \\
\hline
\end{tabular}
\end{table}

$V_0^-$ consists of 1) the e$^-$ kinetic energy, arising from its exchange repulsion from the ``core'' electrons of molecules (atoms), and 2) the energy due to the polarization interaction of e$^-$ with the medium.\footnote{In case of e$^+$ the kinetic contribution to $V_0^+$ is due to the Coulombic repulsion from the nuclei (the exchange repulsion is absent).} According to the theory of the quasifree electron \cite{Spr68}, this polarization interaction may be estimated as a sum of two parts: a) interaction of the e$^-$ with the molecule where it resides, $U_-^{int}$ (to calculate $U_-^{int}$ the electron is considered as an electron cloud smeared over the molecule), and b) interaction of the e$^-$ with all the other molecules, $U_-^{out}=(1-1/\varepsilon)e^2/2R_{\rm WS}$, (this expression is similar to the well-known Born formula for the electron solvation energy).

Experimental values for $V_0^-$ are known for many liquids (Table \ref{V0}). Because of a lack of experimental data on the e$^+$ work functions, we shall admit that they are approximately the same as for e$^-$: $V_0^+ \approx V_0^-$. So we may conclude that $|V_0^+ +V_0^-|\lesssim 1$ eV. Thus $|V_0^+ +V_0^-|$ is less than the variation of the internal energy of the pair, $\approx {\rm Ry}(1-1/\varepsilon^2)/2 \approx 5$ eV, related with the variation in the dielectric screening of the e$^+$-e$^-$ attraction in the bubble formation process.

Note that the use of Eqs. (\ref{2_2}) for the potential energies of the e$^+$ and e$^-$ interaction with the medium, assumes that the polarization interaction remains the same whether e$^+$ and e$^-$ (both in the quasi-free states) are well separated or form the quasi-free Ps atom. Since, for distances larger than the size of a molecule, qf-Ps is nearly an electrically neutral particle, the contributions $U_-^{out} \approx U_+^{out}$, which come from a long-range polarization interaction of the quasi-free e$^+$ and e$^-$ with the medium, should be absent in $U(r_-)+U(r_+)$ in Eq. (\ref{2_1}). Therefore, it is reasonable to consider at least two cases: 1) when the above mentioned polarization correction is neglected and $V_0^- + V_+ \to 0$ and 2) when the terms $U_-^{out}\approx U_+^{out} \approx -1$ eV are subtracted from the work functions and $V_0^- + V_0^+ \to 2$ eV. Both cases are considered below.

\begin{figure}
\centering
 \includegraphics[width=130mm]{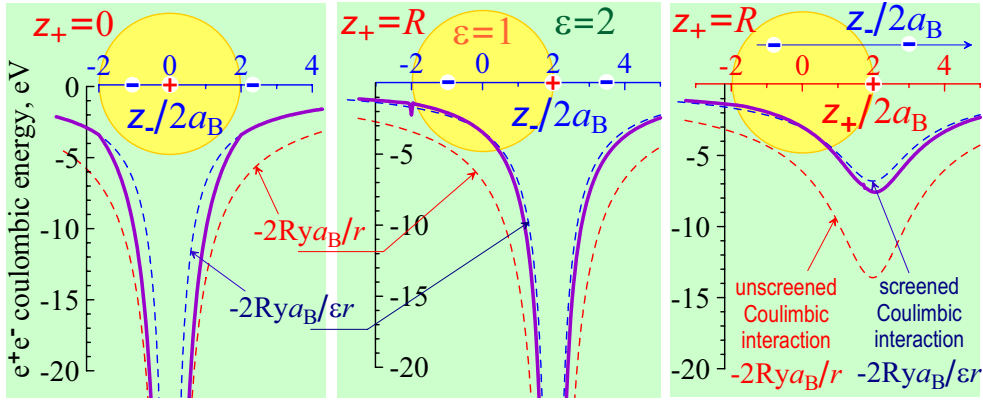}
\caption{Dependence of the e$^+$-e$^-$ Coulombic interaction energy for different locations of e$^+$ and e$^-$ around the bubble (here we adopt the radius of the bubble $R$ equal to $4a_B \approx 2$ \AA). $z_+$ and $z_-$ are the e$^+$ and e$^-$ displacements from the center of the bubble along the $z$-axis. The dashed curves describe the unscreened (red) and completely screened ($\varepsilon$ times less; blue) Coulombic energies between e$^+$ and e$^-$.}\label{BaT_Uc}
\end{figure}

In Eq. (\ref{2_1}) $U_c$ stands for the Coulombic interaction between e$^+$ and  e$^-$ in a polarizable medium. Assuming that the medium has the dielectric permittivity $\varepsilon$ of the bulk and a spherical cavity of radius $R$ (inside the cavity $\varepsilon=1$), one may calculate $U_c$ by solving the Poisson equation. Denoting the e$^+$ and e$^-$ coordinates as ${\bf r}_+$ and ${\bf r}_-$, $U_c$ may be written in the form of the following series via the Legendre polynomials $P_l(x=\cos \theta)$ \cite{Bat02}:
\begin{equation}\label{2_1a}
\frac{U_c(r_+<R,r_-<R)}{\rm Ry} = \frac{2a_B}{r} -\left( 1-\frac{1}{\varepsilon} \right)\frac{2a_B}{R}
 \left(1+\sum_{l=1}^\infty \frac{(1+l)P_l(x)}{1+l+l/\varepsilon} \cdot \frac{r_+^l r_-^l}{R^{2l}}\right);
\end{equation}
$$
\frac{U_c(r_+<R,r_->R)}{\rm Ry} = \frac{2a_B}{\varepsilon r_-}
	 \left( 1+\sum_{l=1}^\infty \frac{(1+2l)P_l(x)}{1+l+l/\varepsilon} \cdot \frac{r_+^l}{r_-^l} \right);
$$
$$
\frac{U_c(r_+>R,r_-<R)}{\rm Ry} = \frac{2a_B}{\varepsilon r} +
	 \left(1-\frac{1}{\varepsilon} \right) \frac{2a_B}{R}
	 \sum_{l=1}^\infty \frac{l P_l(x)}{l+\varepsilon+l\varepsilon)} \cdot \frac{R^{2l}}{r_+^l r_-^l};
$$
$$
\frac{U_c(r_+>R,r_->R)}{\rm Ry} = \frac{2a_B}{\varepsilon r_+}
	 \left( 1+\sum_{l=1}^\infty \frac{(1+2l)P_l(x)}{1+l+l/\varepsilon} \cdot \frac{r_-^l}{r_+^l} \right).
$$
Here, the argument of the Legendre polynomials is $x \equiv \cos \vartheta$, where $\vartheta$ is the angle between the $z$ axis and the direction of ${\bf r}_-$. Note that the summation of these series is simplified considerably when using the following recurrent relationship
$$
 P_l(x) = [(2l-1)xP_{l-1}(x) -(l-1)P_{l-2}(x)]/l.
$$
Particular dependencies of $U_c$ for some selected arrangements of e$^+$ and e$^-$ and the cavity are shown in Fig. \ref{BaT_Uc}. Thus, we are able to take into account the variation of the e$^+$e$^-$ Coulombic energy during the formation of the Ps bubble. Similarly, the dielectric screening is used in the polaron problem and the ion-electron recombination problem (Onsager's formula) \cite{Pek51,Knox63}.

%%%%%%%%%%%%%%%%%%%%%%%%%%%%%%%%%%%%%%%%%%%%%%%%%%%%%%%%%%%
%%%%%%%%%%%%%%%%%%%%%%%%%%%%%%%%%%%%%%%%%%%%%%%%%%%%%%%%%%%%

\subsection{Wave function of the e$^+$e$^-$ pair and minimization of its total energy $\langle H \rangle$}

\begin{figure}[t!]
\centering
 \includegraphics[width=140mm]{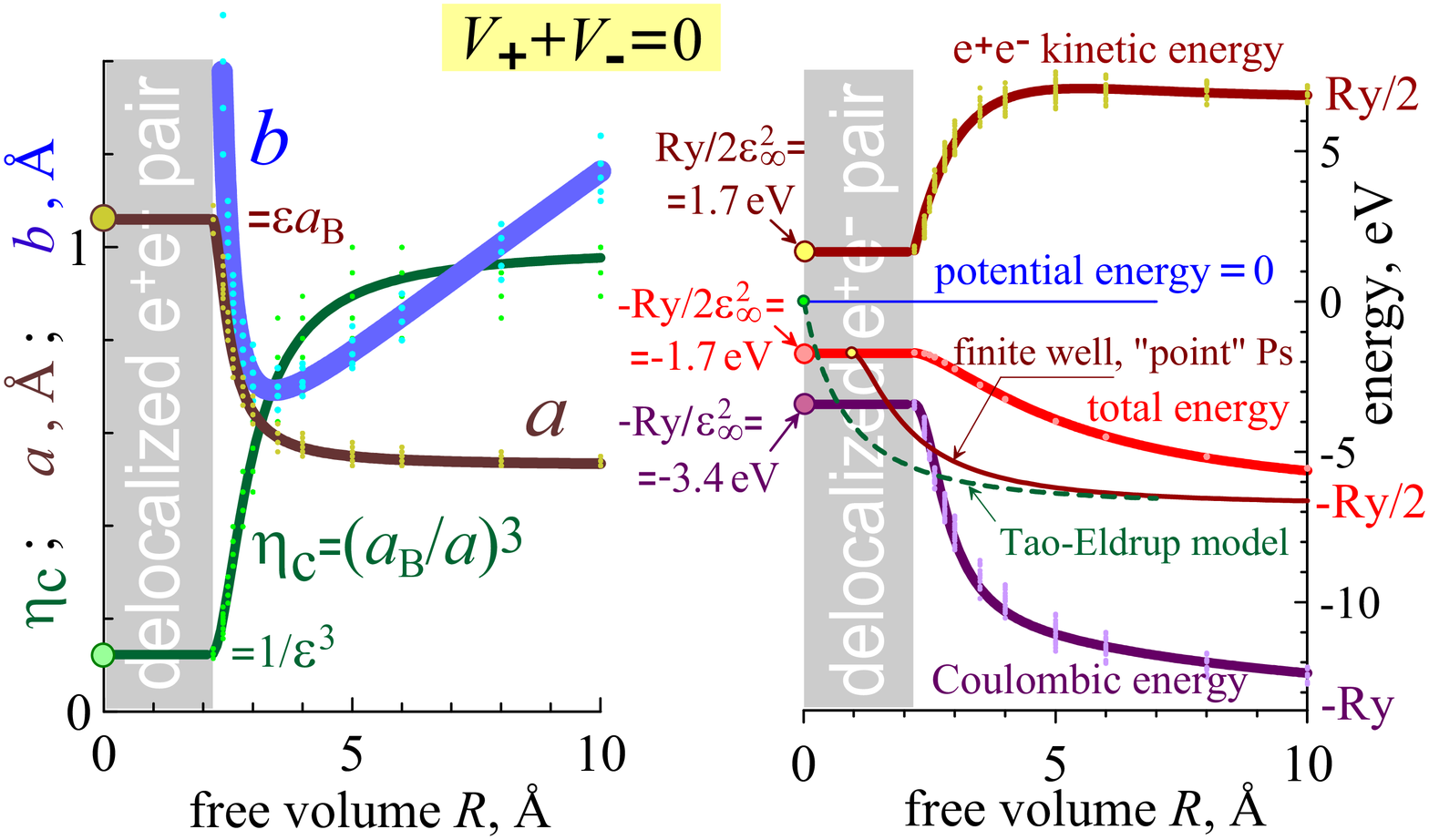}
\bigskip
%\vspace{1mm}
%% \epsfig{file=./V_2_energies.eps, width=140mm}
 \includegraphics[width=140mm]{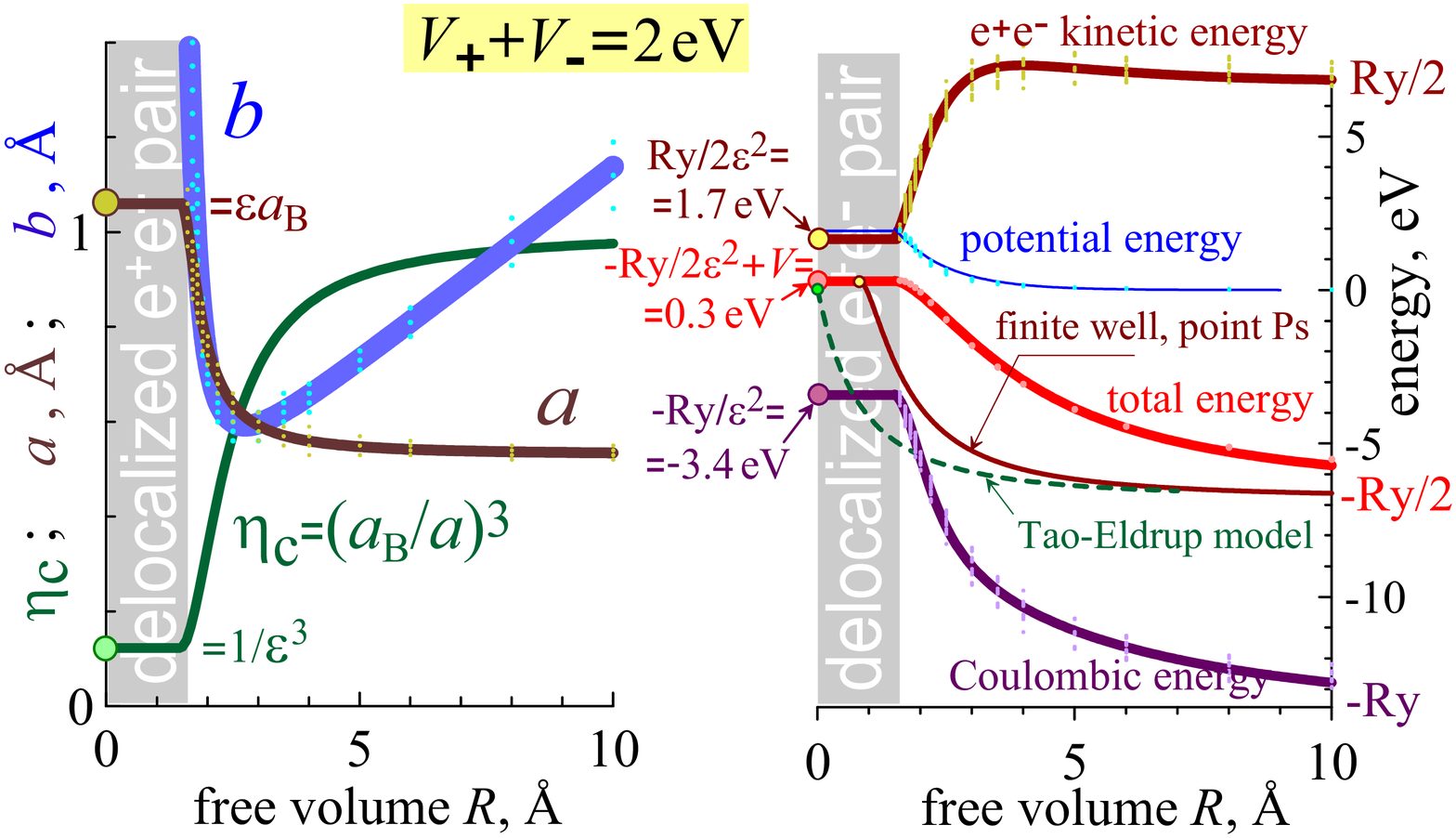}
\caption{Dependencies of the optimal parameters $a$ and $b$ vs. $R$, the bubble radius. They enter the e$^+$e$^-$ wave function and yield the minimum of the total energy $\langle H \rangle$. The relative contact density $\eta_c$ and different energy contributions to $\langle H \rangle$ (at optimal  $a$ and $b$) are shown as well. The upper drawings correspond to the case $V_+ + V_- =0$ and the lower ones to $V_+ + V_- =2$ eV. In both cases it was assumed that $\varepsilon = 2$.}\label{E_R}
\end{figure}

Keeping in mind further use of the variational procedure, let us choose the normalized e$^+$e$^-$ wave function in the following simplest form:
\begin{equation}\label{EM6}
\Psi_{+-}({\bf r}_+, {\bf r}_-) =\frac{\exp(-r/2a -r_{cm}/2b)}{8\pi\sqrt{a^3 b^3}},
\qquad
{\bf r}_{cm}=\frac{{\bf r}_+ + {\bf r}_-}{2},
\quad
{\bf r}={\bf r}_+ - {\bf r}_-.
\end{equation}
In both cases of a rather large bubble and a uniform dielectric continuum, $\Psi_{+-}$ breaks into a product of
two terms: the first one depends on the distance $r$ between e$^+$ and e$^-$, and the second one depends on the
center-of-mass coordinate ${\bf r}_{cm}$.  Parameters $a$ and $b$ are the variational ones, over which we have
minimized the energy of the e$^+$e$^-$ pair:
\begin{equation}\label{2_3}
E(a,b,R) = \langle \Psi_{+-}|H|\Psi_{+-} \rangle \to min ~~~
 \Rightarrow ~~~\hbox{$a(R)$, $b(R)$}.
\end{equation}
The simplest verification of the calculations is to recover two limiting cases. In case of large bubbles ($R\to \infty$), one should reproduce the ``vacuum'' state of the Ps atom: its total energy must tend to $\rm -Ry/2 = -6.8$ eV, the kinetic energy to $\rm +Ry/2$ and the Coulombic energy to $\rm -Ry$. In case of small bubbles ($R\to 0$), the delocalized qf-Ps state must be reproduced. The Schr\"odinger equation for qf-Ps has the same form as for the vacuum Ps, but with the substitution $e^2 \to e^2/\varepsilon$. Then the total qf-Ps energy tends to $V_0^+ +V_0^- \rm -Ry/2\varepsilon^2$, its kinetic part tends to $\rm +Ry/2\varepsilon^2=1.7$ eV ($\varepsilon = 2$) and the Coulombic energy tends to $\rm -Ry/\varepsilon^2=-3.4$ eV. Fig. \ref{E_R} displays optimal values of $a$ and $b$ as well as different contributions to the total energy of the e$^+$e$^-$ pair when $V_0^+ + V_0^- =0$ and 2 eV.

%%%%%%%%%%%%%%%%%%%%%%%%%%%%%%%%%%%%%%%%%%%%%%%%%%%%%%%%%%%
%%%%%%%%%%%%%%%%%%%%%%%%%%%%%%%%%%%%%%%%%%%%%%%%%%%%%%%%%%%%

\subsection{Relative contact density and pick-off annihilation rate}
\begin{figure}[t!]
\centering
\begin{minipage}[h]{70mm}
 \includegraphics[width=70mm]{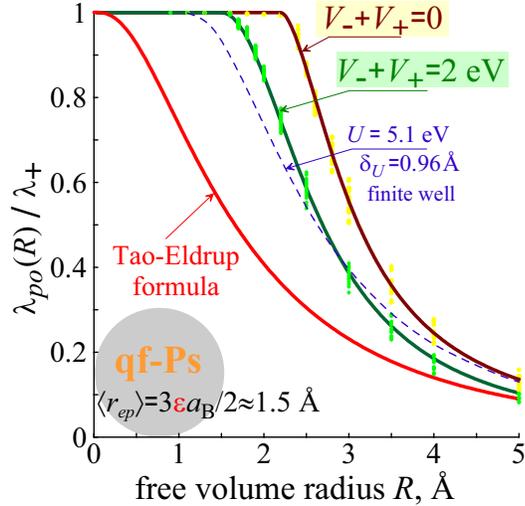}
\end{minipage}
~~~
\begin{minipage}[h]{90mm}
\caption{Pick-off annihilation rate constant of Ps, localized in a bubble of $R$ when
$V_+ + V_- =0$ (green curve) and $V_+ + V_- = 2$ eV (brown curve). For small $R$ ($\lesssim 2$ \AA)
the calculated values of $\lambda_{po}$ are equal to $\lambda_+$. The red line shows pick-off annihilation
rate constant, calculated according to the Tao-Eldrup formula. The dashed line is the calculation according to
the finite potential well model (for comparison we adopted that the depth of the well is
$(1-1/\varepsilon^2){\rm Ry}/2 \approx 5.1$ eV and its radius is $R$. The minimal radius of the well
when there appears an energy level is $\delta_U= 0.96$ \AA).}\label{L_po}
\end{minipage}
\end{figure}

In the framework of the developed scheme using the wave function (\ref{EM6}) it is easy to obtain
the relative contact density $\eta_c$ in Ps atom:
\begin{equation}\label{B7_2}
\eta_c = \frac{\int \int d^3 {\bf r_+} d^3 {\bf r_-} |\Psi      _{+-}({\bf r}_+, {\bf r}_-)|^2 \delta({\bf r}_+ - {\bf r}_-)}
			  {\int \int d^3 {\bf r_+} d^3 {\bf r_-} |\Psi^{vac}_{+-}({\bf r}_+, {\bf r}_-)|^2 \delta({\bf r}_+ - {\bf r}_-)}
	   = \frac{a^3_B}{a^3(R)}.
\end{equation}
This quantity determines the observable Ps annihilation rate constant (including the case with applied permanent magnetic field). The resulting dependencies of $\eta_c$ are shown in Fig. \ref{E_R} (on the left). Because, for qf-Ps,  parameter $a$ is equal to $\varepsilon a_B$, for qf-Ps the value of $\eta_c$ should be $1/\varepsilon^3 =1/8$, which is well recovered in numerical calculations. When $R$ increases, $\eta_c$ approaches unity, because $a$ tends to its vacuum value $a_B$. Knowing the expression for the wave function (\ref{EM6}), one may calculate the positron overlapping $P_R$ with molecular electrons, surrounding the Ps atom, and therefore find out the pick-off annihilation rate constant:
\begin{equation}\label{1_67a}
\lambda_{po}(R) \approx \lambda_+ P_R, \qquad
 P_R \approx \int_{r_+>R} d^3 {\bf r}_+ \int d^3 {\bf r}_- \left| \Psi_{+-}  ({\bf r_+} ,{\bf r_-}) \right|^2 .
\end{equation}
Here, $\lambda_+ \approx 2$ ns$^{-1}$ is the annihilation rate constant of ``free'' positrons. Results of calculations of $\lambda_{po}(R)$ for optimal $a$ and $b$ values, which correspond to the minimal Ps energy at a given $R$, are shown in Fig. \ref{L_po}.

%%%%%%%%%%%%%%%%%%%%%%%%%%%%%%%%%%%%%%%%%%%%%%%%%%%%%%%%%%%
%%%%%%%%%%%%%%%%%%%%%%%%%%%%%%%%%%%%%%%%%%%%%%%%%%%%%%%%%%%%

\subsection{Results and discussion}

1) It is usually considered that Ps is a solvophobic particle, i.e., it forms a bubble when entering a liquid because of exchange repulsion between e$^-$ in Ps and the surrounding molecular electrons. If the work functions of e$^+$ and e$^-$ are negative ($V_0^+ \approx V_0^- < 0$), each particle considers a cavity as a potential barrier. So they are pulled to the bulk by polarization interaction with the medium. Nevertheless, even in this case the Ps bubble may be formed due to an enhancement of the Coulombic e$^+$e$^-$ attraction inside the cavity (no dielectric screening inside). This feature cannot be taken into account when Ps is simulated as a point particle.

2) It is seen that the behavior of the total energy of the pair (red curves in Fig. \ref{E_R}) strongly differs from the Tao-Eldrup prediction (green dashed curves; the first term in Eq. (\ref{9_2}), where $R_\infty$ is replaced by $R$), as well as from the expectation based on the finite potential well model (brown curves in Fig. \ref{E_R}; the Coulombic potential cannot be approximated well by a rectangular spherically symmetric potential). The same is true for the pick-off annihilation rate, Fig. \ref{L_po}.

3) Calculations demonstrate one common feature: up to $R\lesssim 1.5-2.2$ \AA\ all dependencies remain the same as in a medium without any cavity, but at larger $R$ there are significant deviations. This is related to the known quantum mechanical phenomenon -- absence of a bound state of a particle in a small finite 3d-potential well. In such cavities, Ps cannot be bound, it does not exert any repulsive pressure on their walls and does not stimulate their transformation towards the equilibrium Ps bubble. The possibility of finding a suitable preexisting cavity, sufficient at least for preliminary localization of qf-Ps, may be a limiting factor for the formation of the Ps bubble state.

4) One may find an equilibrium Ps bubble radius by minimizing the sum of the total e$^+$e$^-$ energy $\langle H \rangle$ and the surface energy of the bubble. For water it turns out to be 5-5.2 \AA\ which is about 2 \AA\ larger than predicted by the Tao-Eldrup model. For such a large bubble, the relative contact density is $\eta_c\approx 0.9$, Fig. \ref{E_R}. It is somewhat higher than the experimental values (0.65-0.75 \cite{Ste11}). This discrepancy may indicate that e$^+$ and e$^-$ really interact with a medium in a different way, for example, $V_0^- > V_0^+$. It means that the Ps electron may be trapped by a cavity, and e$^+$ will be bound to this trapped e$^-$ by the Coulombic attraction. This scenario may be also considered in the framework of the present approach, but the expression for the trial wave function of the new pair must be written in an ``asymmetric'' (towards e$^+$ and e$^-$) form:
\begin{equation}
\Psi_{+-}({\bf r}_+, {\bf r}_-) \approx \frac{\exp(-|{\bf r}_+ - {\bf r}_-|/2a  -r_-/2b)} {8\pi\sqrt{a^3 b^3}}.
\end{equation}

5) Any Ps bubble model reduces the original many-body (multi-particle) problem to a simpler one, that of one or two particles in an external field, which simulates the interaction with the medium. To calculate this field one usually relies on some macroscopic approaches. However, their validity always remains uncertain (for example, how to relate the actual arrangement of molecules around the Ps bubble with the jump of dielectric permittivity outside the bubble and so on).

\medskip

This work is supported by the Russian Foundation of Basic Research (grant 11-03-01066). 
%and Federal Agency on Atomic Energy.

\end{document}